# Thermal tuning capabilities of semiconductor metasurface resonators

Tomer Lewi[1,2*], Nikita A. Butakov[1] and Jon A. Schuller[1]

[1] Department of Electrical and Computer Engineering, University of California, Santa Barbara, California 93106, USA
[2] Faculty of Engineering and Bar-Ilan Institute for Nanotechnology and Advanced Materials (BINA), Bar-Ilan University, Ramat-Gan 5290002, Israel
*E-mail: tomer.lewi@biu.ac.il

**Abstract**

Metasurfaces exploit optical phase, amplitude and polarization engineering at subwavelength dimensions to achieve unprecedented control of light. The realization of all dielectric metasurfaces has led to low-loss flat optical elements with functionalities that cannot be achieved with metal elements. However, to reach their ultimate potential, metasurfaces must move beyond static operation and incorporate active tunability and reconfigurable functions. The central challenge is achieving large tunability in subwavelength resonator elements, which requires large optical effects in response to external stimuli. Here we study the thermal tunability of high-index silicon and germanium semiconductor resonators over a large temperature range. We demonstrate thermal tuning of Mie resonances due to the normal positive thermo-optic effect (dn/dT >0) over a wide infrared range. We show that at higher temperatures and longer wavelengths, the sign of the thermo-optic coefficient is reversed, culminating in a negative induced index due to thermal excitation of free carriers. We also demonstrate the tuning of high order Mie resonances by several linewidths with a temperature swing of ΔT<100K. Finally, we exploit the large NIR thermo-optic coefficient in Si metasurfaces to realize optical switching and tunable metafilters.

**Introduction**

Metasurfaces are planar optical structures composed of ordered subwavelength resonators, designed to manipulate light through arbitrary wavefront shaping [1] [2] [3] [4]. Recently, this field has witnessed tremendous progress by adopting an all-dielectric approach [2], giving rise to several dielectric metasurface demonstrations including achromatic and broadband metalenses [5] [6] [7], axicon lenses[8], sub-diffraction focusing[9], nonlinear generation[10] [11] [12] beam deflectors[8,13] [14], wave plates and beam converters [15] [16] [17] [18], holograms[19] [20] [21], antireflection coatings[22] and magnetic mirrors[23], to name a few. So far, however, most metasurfaces are implemented for static operation and optimized for limited bandwidth of operation. To reach the next level of dynamic control over light, metasurfaces must include active and reconfigurable functionality that will drastically increase their potential and unlock a vast array of new application possibilities.

The fundamental challenge for achieving reconfigurable operation is to obtain large and continuous modulation of optical properties within subwavelength and low-Q meta-atom resonators [24,25]. Desirable tuning mechanisms continuously shift the resonance frequency of the metastructure with at least one linewidth of maximal shift, thus enabling significant modulation of both amplitude and phase. These challenges have motivated several studies exploring different approaches, designs, and materials that provide extreme tunability. Previous investigations of active tuning in dielectric metasurfaces and meta-atoms have focused on ultrafast free-carrier injection[26][27][28][29][30], coupling to liquid crystals[31][32], to atomic vapor[33] or to ENZ materials[34,35], phase change materials[36][37] and MEMS[38][39][40]. However, none of these approaches provide a viable solution for a fully reconfigurable metadevice where at each subwavelength meta-atom the phase and amplitude can be individually and continuously tuned to provide an arbitrary phase profile. Free-carrier approaches have only shown sub-linewidth resonance tuning, along with severe losses, as they require exceptionally high carrier generation rates. Phase change materials only allow non-continuous switching between two distinct states while coupling to background media and MEMS based approaches do not provide direct modulation of the resonator itself but rely on mechanical actuation or index modulation of substrates. Recent studies have showed that the thermo-optic effect (TOE), i.e. refractive index variation with temperature dn/dT can be used to induce large (Δn~1) and continuous index shifts in materials having extraordinary thermal dependence[41][42]. Recently, the TOE was also used to actively tune Si metasurfaces, but only at a limited temperature range (273-573K)[43]. Here, we present a thorough study of the thermal tunability of Ge and Si meta-atoms and metasurfaces over a large temperature range 80-900K. We demonstrate thermal tuning of Mie resonances due to the normal positive thermo-optic coefficient (TOC) (dn/dT >0) over a wide infrared range. At higher temperatures and longer wavelengths, we show that thermal excitation of free carriers (FC) becomes significant due to the bandgap shrinkage of the semiconductors, causing a reduction in dn/dT. With further increased temperatures, the sign of TOC is reversed (dn/dT<0) culminating in a total negative induced index. We also demonstrate the tuning of high order Mie resonances by several resonance linewidths with a temperature swing of ΔT<100K. Finally, we exploit the larger TOC of Si at NIR wavelengths for realizing amplitude modulators and tunable metafilters with Si metasurfaces.

Thermo-optic (TO) effects provide an ideal test bed for demonstrating and elucidating reconfigurable metasurface properties. TO tuning can provide large index shifts with no added losses and be integrated into electrically-controlled architectures[44,45]. Thus, TO tunability forms the basis for many reconfigurable integrated photonic devices [44,45,46,47]. However, the TOC of most materials is small for subwavelength applications, hence typical TO applications exploit small index changes acting over distances much larger than a wavelength to achieve useful modulation. For efficient modulation of subwavelength resonators, the maximally induced index shift Δn should tune the resonance wavelength by more than its linewidth (Δλ/FWHM>1, where Δλ is the resonance wavelength shift and FWHM is the full width at half max of the linewidth). The route for achieving this tunability is by maximizing the TOE using extraordinary materials[41,42,48,49] and/or narrowing the resonance linewidth using high-Q modes[41] such as supported by asymmetric[32] or fano-resonant[10] metasurfaces or originating from bound states in the continuum[50]. Here, we study the TO tuning capabilities of Si and Ge – the most commonly used materials for dielectric metasurfaces and nanophotonics. The TO coefficients of these semiconductors are amongst the highest of natural materials[51] which, along with their high refractive indices and CMOS compatibility, makes them very attractive materials for reconfigurable metasurfaces. However, the typical TOC values (~ 1 – 5 x $10^{-4}$ $K^{-1}$) requires large temperature modulation which may cause problems if the TOC is strongly temperature dependent[41,42]. In the mid-infrared (MIR) range, for instance, working at high temperatures can generate FC densities in semiconductors that dramatically alter the optical constants due to Dude-like dispersion. The total induced index shift of the semiconductor due to a *positive* temperature gradient is the sum of contributions from the normal TO effect and the thermal FC effect: Δn=Δn$_{TO}$+Δn$_{FC}$. In vast majority of materials, the TOC is positive hence Δn$_{TO}$>0 while the FC term has a negative contribution $\Delta n_{FC}<0$ (due to the plasma frequency blue shift caused by FCs, in a Drude model). The thermal FC term ($\Delta n_{FC}$) is particularly strong for low bandgap semiconductors with small effective masses[24][41][42] (supporting information), but in most semiconductors is negligible below ~500K.

**Results and Discussion**

The induced index shift of Si and Ge as a function of wavelength (2-16µm) and temperature (80-850K) is presented in Figures 1a and 1b. Refractive index shifts Δn are calculated with respect to index values at room temperature (RT). The index shifts in Si are almost wavelength independent for ~ T<700K, where the normal positive dn/dT is responsible for a near linear increase in Δn with temperature. For T>700K, Δn is wavelength dispersive since the density of thermally generated free carriers is no longer negligible (supporting information Figures S1 and S2). Extracted dn/dT linecuts at different temperatures (Figure 1c), illustrate the temperature dependence and chromatic dispersion of dn/dT at various temperature regimes, specifically when dn/dT switches signs, i.e. dn/dT<0, for T=800K (orange line). The strong dn/dT wavelength dispersion at 800K (orange line) is due to a 15% decrease in bandgap which generates an intrinsic free carrier density of $n_i$~$10^{17}$ cm$^{-3}$ (supporting information Figures S1 and S2) and associated negative $\Delta n_{FC}$. Thus, there is a point where both effects perfectly cancel out $\Delta n_{TO}=-\Delta n_{FC}$ and the total Δn is zero. In Ge, these thermal free carrier effects are further accentuated, as seen in Figures 1b and 1d. The smaller bandgap of Ge (0.66eV at RT) compared to Si (1.12eV at RT) and the lower FC effective mass, causes a stronger FC contribution for the same temperature gradients. Therefore FC effects play a significant role even at 500K. For more elevated temperatures and longer wavelengths FC effects dominate over normal TO effects, leading to a negative and highly dispersive dn/dT at MIR and LWIR wavelengths. For example, at T=800K (orange line), dn/dT is almost an order of magnitude larger in magnitude at λ=14µm (-18 x 10$^{-4}$ K$^{-1}$) compared to λ=2µm (2 x 10$^{-4}$ K$^{-1}$).

To investigate thermal tunability capabilities, we study Si and Ge single spherical meta-atom resonators fabricated by laser ablation[24,52] . Examples of Si (r=1.9µm) and Ge (r=1.54µm) meta-atom Mie resonators are illustrated in Figures 2a and 2b. A series of multipolar Mie resonances in the range 4-14µm are observed using both analytical calculations (red dashed), FDTD simulations (red), and single particle infrared microspectroscopy (black). These multipolar resonances are labeled (see panels 2c and 2d) according to their polarization (Magnetic or Electric) and mode order (Dipole, Quadrupole, Hexapole, etc'). The temperature dependent spectra of these resonators (80-873K for Si and 123-773K for Ge, see experimental section for more details) are presented in Figures 2c and 2d. Spectral shifts are observed for all resonances as a response to the thermal modulation of the refractive index of the resonators.

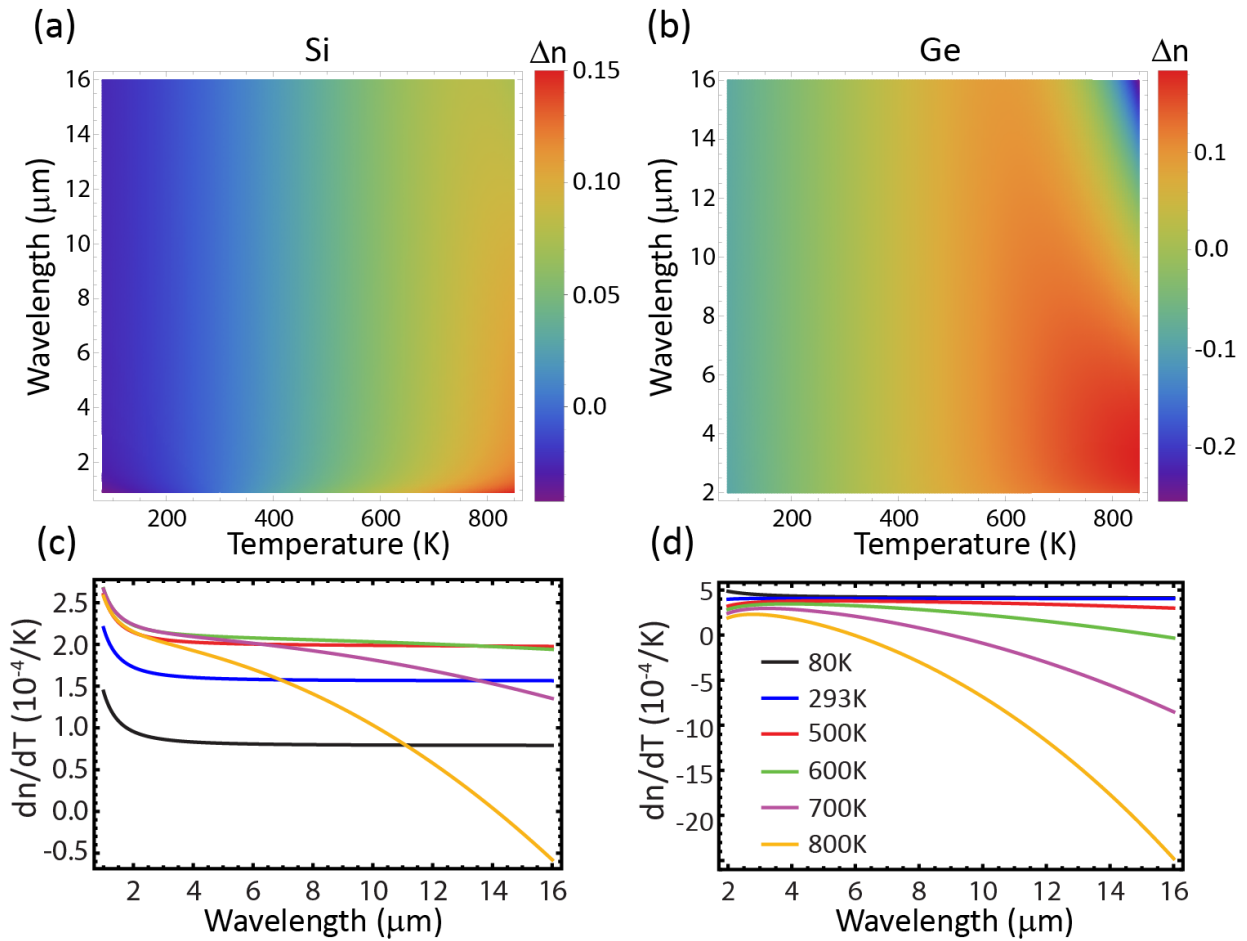

Figure 1: Thermally induced index shifts in Si (a) and Ge (b) as a function of wavelength (2-16µm) and temperature (80-850K). (c) &(d) The chromatic dispersion of dn/dT at various temperatures in Si (c) and Ge (d). Each color plot corresponds to dn/dT at different temperature as detailed in the legend of (d).

When thermally generated free carriers are negligible, all resonances are expected to red shift in response to a positive thermal gradient due to the normal positive dn/dT in both Si and Ge. This behavior is confirmed for both particles between cryogenic temperatures and up to 500K. For more elevated temperatures the shift in the long wavelength dipole modes deviates from the near-linear increase in wavelength. This behavior is more prominent in Ge due to its smaller bandgap and lighter effective mass (supporting information section 1), which respectively facilitate higher free carrier concentrations at each given temperature and a larger index shift for a given carrier density. In Figures 2e and 2f we track the temperature dependent resonance wavelength shifts of the MD and ED, with respect to the RT resonance wavelength, and extract the corresponding index shifts Δn. Using our model for the temperature dependent permittivity of the semiconductors (supporting information section 1), we compare experiments to calculated resonance and index shifts of the ED and MD modes (also see Figure S3, supporting information). For Si, the curvature of dn/dT significantly changes only above ~770K and exclusively for the MD mode (due to its longer wavelength), as seen by the peak in both Δn and Δλ at~800K. Above the peak, the induced index and resonance wavelength decrease due to the generation of $n_i>10^{17}cm^3$ free carriers; the sign of dn/dT has reversed (dn/dT <0) and now causes a slight blue shift for the MD mode (at this point the curvature of the ED mode is also affected as it starts to flatten).

For the Ge resonator (Figure 2f), FC effects are more prominent and emerge at lower temperatures. Due to its smaller bandgap and lower effective masses, the dispersion of the induced index Δn of the MD mode starts to flatten at ~550K, peaks at 600K then decreases at higher temperatures (dn/dT<0). For even more elevated temperatures free carriers dominate the index change giving rise to a larger magnitude|dn/dT|. The behavior of the ED is similar, but the change in curvature of Δn is shifted to higher temperature T~675K since the FC effect is smaller at shorter wavelengths. Altogether, these results show that the dispersion, the sign, and magnitude of dn/dT can be controlled with temperature in low and moderate bandgap semiconductor resonators through the generation of free carriers. This can be used to engineer the dispersion of dn/dT and for tuning of infrared meta-atoms and resonators.

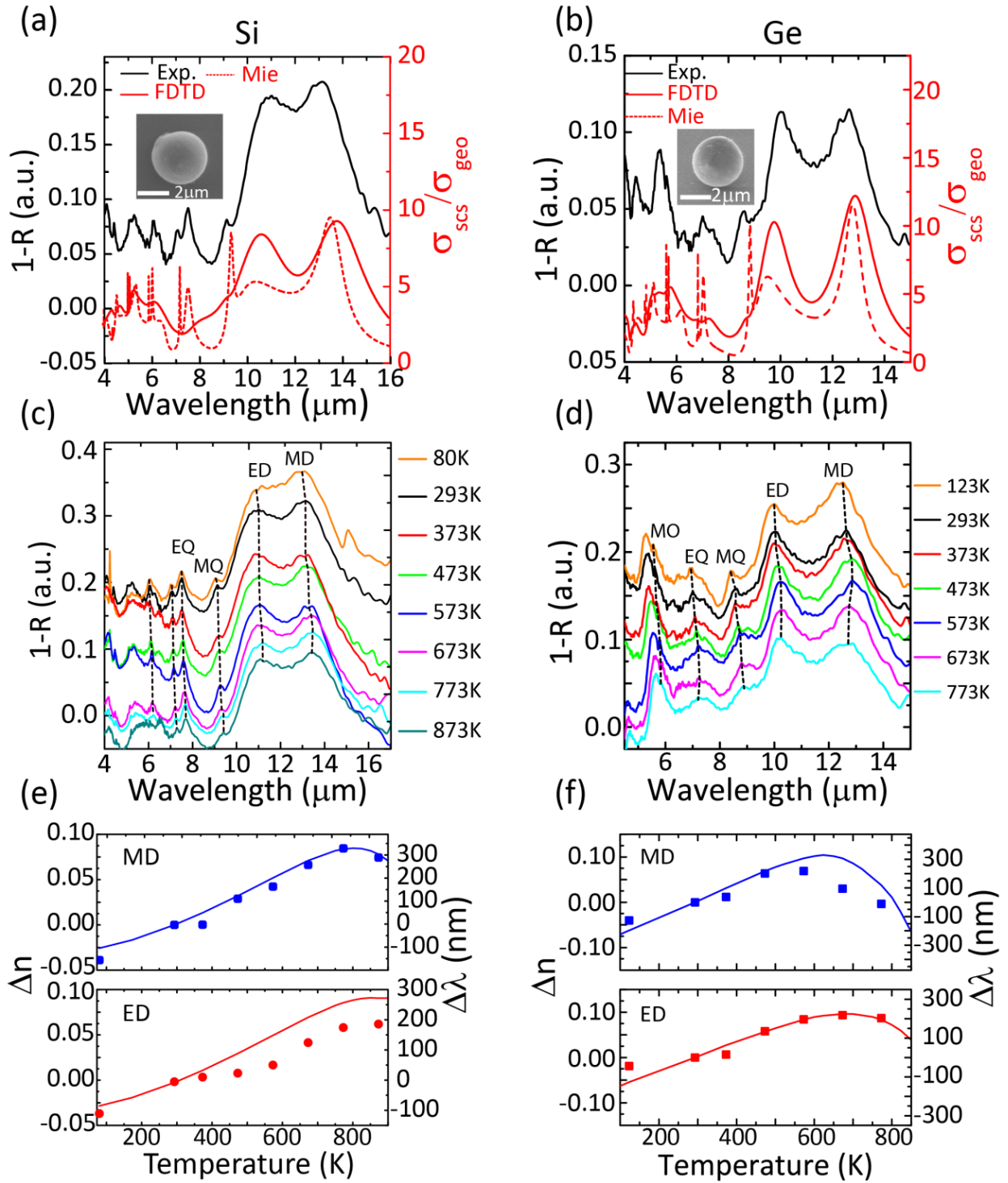

Figure 2: Thermal tunability in single semiconductor meta-atom resonators. (a) & (b) Infrared spectra of Si ((a), r=1.9µm) and Ge ((b), r=1.54µm) spherical Mie resonators. Experimental spectra show good agreement with the calculated Mie scattering and FDTD cross-sections $\sigma_{sca}$ normalized to the geometric cross-section $\sigma_{geo}$. A series of multipolar resonance peaks are visible in the spectra and correspond to dipole, quadrupole, hexapole, octapole, etc. (c) & (d) Temperature dependent spectra of the spherical resonators between 80K and 900K. The spectral shifts of magnetic and electric dipole (MD and ED) and quadrupole (MQ and EQ)

modes are highlighted. The dashed lines track the temperature dependent resonance peaks and are a guide to the eye. The magnetic octapole (MO) shift is also highlighted in the Ge resonator and is elaborated on in Figure 3. (e) & (f): The extracted induced index and resonance wavelength shifts as a function of temperature for MD and ED modes in Si (e) and Ge (f). Dots are experimentally extracted values while solid lines are calculated shifts based on Mie theory.

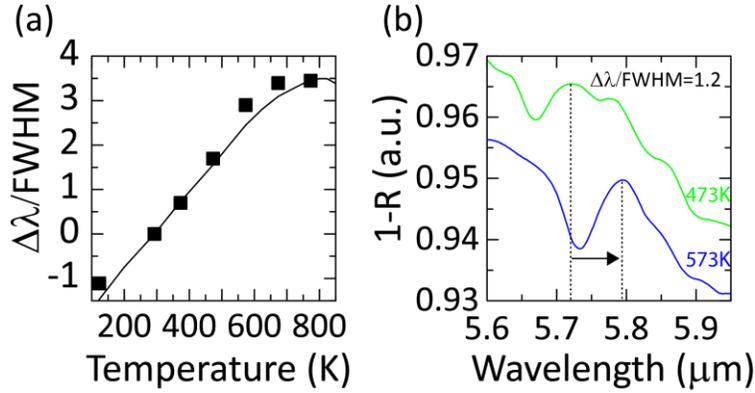

Figure 3: Normalized tunability (Δλ/FWHM) of high-order MO mode in the Ge resonator. (a) Temperature dependent MO shifts normalized to the resonance linewidth. (b) Tuning by more than one resonance linewidth (Δλ/FWHM=1.2) with a ΔT=100K temperature gradient.

The ability to maximize the control over the phase and amplitude of incident light depends on the capability of tuning resonance wavelengths by more than one linewidth Δλ/FWHM>1. Figure 3 presents the normalized tunability (Δλ/FWHM) of a high order magnetic octapole mode (MO) that is tuned by ~ 4.5 normalized linewidths across the 123-773K temperature range. This wide normalized tunability becomes possible due to an order of magnitude increase in the resonance Q-factor of $Q_{MO}$=88.1 compared with $Q_{MD}$=8.2 for the MD mode. As evident from Figure 3a, the normalized tunability exhibits a linear like dependence up to 675K, with almost one normalized linewidth per 100K. For ~ T>675K, the linear dependence breaks as free carrier effects become significant which leads to a rapid saturation of tunability with increased temperature. Figure 3b, presents the spectral shift of Δλ/FWHM=1.2 with a temperature gradient of ΔT=100K. Such tunability with a practical temperature difference of ΔT=100K, would be useful for implementing reconfigurable high-Q Ge metasurfaces.

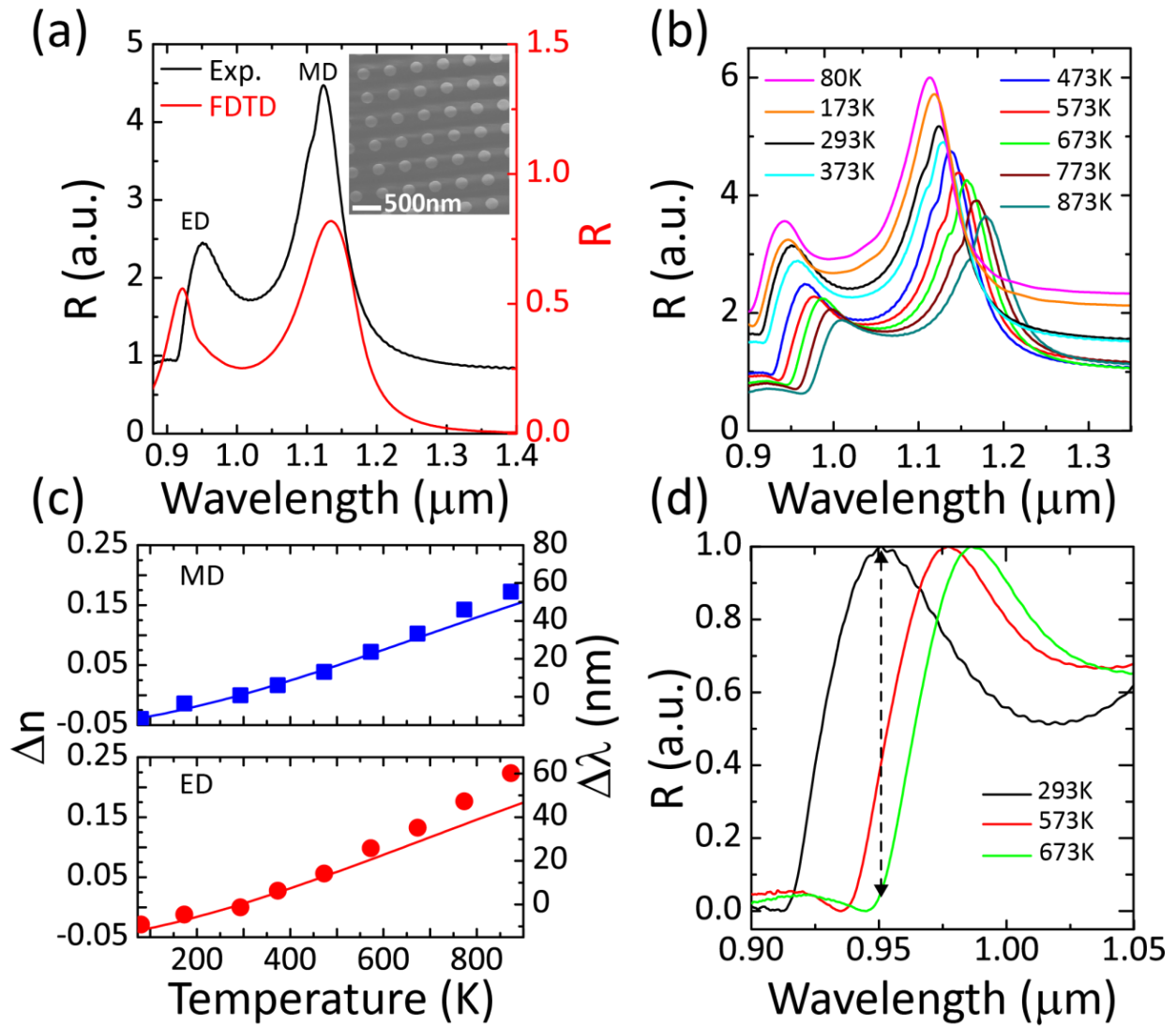

Figure 4: Thermally reconfigurable Si metasurfaces. (a) RT NIR reflection spectra of a Si metasurface disk array with disk diameter and height of d=290nm and h=280nm respectively and periodicity a=590nm on a SiO2 substrate. Experimentally measured (black) and FDTD (red) spectra are in good match showing the fundamental MD and ED modes. The inset presents an SEM image of the metasurface. (b) Temperature-dependent spectra exhibiting continuously red-shifted MD and ED resonances with increasing temperature. The spectra are vertically shifted along the y-axis for visibility. (c) Thermally induced index and wavelength shifts of MD and ED modes. (d) Thermally reconfigurable metafilter demonstration, exhibiting 4 dB and 13dB amplitude modulation when a temperature gradient of ΔT=280K and ΔT=380K, is applied, respectively.

A different route for increasing the TO normalized tunability is to exploit the larger TOC at shorter wavelengths, in the vicinity of the material bandgap. As seen in Figure 1c, the dn/dT values are up to 50% larger in Si around 1μm. Below, we demonstrate thermally reconfigurable metasurfaces in near infrared (NIR) range using a Si disk array on SiO$_2$ substrate. Figure 4a shows experimental and FDTD reflection spectra of a silicon disk array with disk diameter and height of d=290nm and h=280nm respectively and periodicity a=590nm. Both experimental and FDTD spectra show pronounced MD and ED resonances (Q~21) at λ=1.12µm and λ=0.95 µm, respectively. Experimental temperature dependence of these resonances is presented in Figure 4b (the spectra are vertically shifted along the y-axis for visibility), where both dipole modes are continuously red shifted with increasing temperatures. The extracted index and resonance wavelengths shifts Δn and Δλ are presented in figure 4c for both dipole modes (Δn and Δλ are calculated with respect to index and resonance wavelengths values at RT). The higher TOC at shorter wavelengths is evinced by a larger induced index shift Δn for the ED mode compared to the MD mode. Extracted dn/dT values for both MD and ED modes are similar or slightly higher than reported values [51]. These high dn/dT values at the ED resonance wavelengths, combined with relatively narrow linewidths (Q ~21), allows for tunable metafilter operation (Figure 4d) with amplitude modulation of more than 95% (13dB) for ΔT=380K at λ=950nm (60% (4dB) amplitude modulation for ΔT=280K). Further reduction of the required temperature modulation may be achieved by engineering ultrahigh-Q metasurfaces (such as the high order Mie resonance presented in Figure 3) that enable normalized tunability Δλ/FWHM >1 with temperature modulation of tens of kelvin or less[41][32] . Importantly, thermal tunability in Si metasurfaces can be extended to the visible range, with improved performance due to the expected increase of dn/dT at shorter wavelengths. Also, thermal tuning can be implemented using electrically-controlled architectures with independent control for each element[45,48].

**Conclusions**

In summary, we studied the thermal tuning capabilities of Si and Ge single resonators and metasurfaces over the large temperature range 80-873K. We demonstrated temperature-dependent resonance frequency shifts that follow a modified model of the traditional TOE that takes into account effects of thermally generated free carriers. We showed that at low and intermediate temperatures, all resonances red-shift according to the normal positive dispersion of TOC (dn/dT>0). At higher temperatures and longer wavelengths however, thermally generated free carriers contribute a negative component to the total TOC and can reverse its sign, i.e. yielding a dn/dT<0. This dispersion anomaly is evinced by a continuous change in the resonance shift from red to blue-shift. We also demonstrated more than a unit of normalized tunability Δλ/FWHM =1.2 of high order octapole modes in Ge resonators with a temperature gradient of ΔT<100K. Ultimately, we exploit the larger TOC at short NIR wavelengths in Si to demonstrate thermally reconfigurable metasurface functionality such as amplitude modulators and tunable metafilters. This work highlights the opportunities and potential of thermally tunable semiconductor metasurfaces and can pave the way to efficient high-Q reconfigurable metadevices.

**Methods**

*Si and Ge* spherical resonators were fabricated by femtosecond laser ablation. In these experiments, we used a commercial femtosecond laser system (Spitfire, Spectra Physics) delivering ~ 1 mJ pulses with ~120 fs duration with central wavelength of 800 nm and variable repetition rate. Pulse energies ranging between 20 and 200 μJ at 20 Hz repetition rate were used in ablation experiments. More details on ablation experiments can be found in previous works [24,52].

Si metasurfaces were fabricated per a previously reported procedure[53]: Amorphous Silicon was deposited onto fused Quartz substrates with an Advanced Vacuum PECVD. Patterning was done using ultraviolet photolithography and deep reactive-ion-etching process using a PlasmaTherm 770 SLR system.

*Optical Characterization:* single particle spectroscopy at various temperatures was conducted using an FTIR (Vertex 70, Bruker) coupled to an infrared microscope (Hyperion 3000, Bruker) using a thermal stage (THMS600, Linkam). More details on single particle spectroscopy were reported elsewhere[24]. Finite difference time domain (FDTD) calculations were performed using the Lumerical Solutions FDTD Solver, Version 8.7.3. A non-uniformal conformal mesh was used. A mesh size at least 10x smaller than the minimum wavelength in the material was used with boundary conditions of perfectly matched layers.

**Conflicts of interest**
There are no conflicts to declare

**Acknowledgments:**

This work was supported by the Air Force Office of Scientific Research (FA9550-16-1-0393) and by the UC Office of the President Multi-campus Research Programs and Initiatives (MR-15-328528). Microscopy was performed with support from MRSEC Program of the NSF under Award No. DMR 1121053; a member of the NSF-funded Materials Research Facilities Network. Numerical calculations were performed with the support from the Centre for Scientific Computing from the CNSI, MRL: an NSF MRSEC (DMR-1121053), NSF CNS-0960316. N.A.B. acknowledges support from the Department of Defense NDSEG fellowship.

# Supporting information for: "Thermal tuning capabilities of semiconductor metasurface resonators"

*Tomer Lewi*, Nikita A. Butakov and Jon A. Schuller*

1. **Thermo-optic coefficient**

The thermo-optic (TO) coefficient in a transparent spectral range can be defined as[1]:

$$2n\frac{dn}{dT} = (n_0^2 - 1)\left(-3\alpha R - \frac{1}{E_{eg}}\frac{dE_{eg}}{dT}R^2\right) \quad (1)$$

where n, $n_0$ and T are the refractive index, the low frequency refractive index and temperature respectively, α is the linear thermal expansion coefficient, and $R = \frac{\lambda^2}{\lambda^2 - \lambda_{ig}^2}$ where $\lambda_{ig}$ is the wavelength corresponding to the temperature-invariant isentropic bandgap[1], and $E_{eg}$ is the temperature-dependent excitonic bandgap. The temperature variation of the excitonic bandgap $E_{eg}(T)$ is typically the dominant contribution to the TO coefficient. For simplicity, Eq. 2 can be rewritten as:

$$\frac{dn}{dT} = \frac{1}{2n}(GR + HR^2) \quad (2)$$

Where $G = -3\alpha(n_0^2 - 1)$ and $H = -\frac{1}{E_{eg}}\frac{dE_{eg}}{dT}(n_0^2 - 1)$. It is clear that in the normal spectral regime (λ>$\lambda_{ig}$ and therefore R>0) the sign and value of the TO coefficient is determined by values of G and H. The contribution from G is usually negative because the thermal expansion coefficient α is positive for most materials, which dictates that G is negative. Also, the contribution of G is usually smaller than H since α is small (~$10^{-6}$/°C). Typically, the temperature variation of the excitonic bandgap $dE_{eg}/dT$ is large (~$10^{-4}$eV/°C) and negative (similarly to the bandgap energy $dE_g/dT$ which is also negative for most materials), that is as temperature increases the bandgap decreases. Since the first factor in H is also negative (-1/$E_{eg}$) it follows that H is usually positive and is the dominant contribution to the TO coefficient. Indeed this is the case for the vast majority of semiconductors as can also be seen in Figure 1 of main text where Si and Ge TOC are always positive (dn/dT>0) for temperatures T<600K.

Specifically for Si and Ge, large number of *n(T,λ)* at different temperatures and wavelengths are available in literature which made it possible to extract and express G(T) and H(T) as a quadratic function of temperature. For the Ge resonators in this work, we used the following relations in order to model the normal TO coefficient and *n(T)* of Ge[1]:

$$G = 0.324 - 2.092 * 10^{-2}T + 5.0398 * 10^{-5}T^2 - 4.1434 * 10^{-8}T^3 \tag{3}$$

$$H = 14.336 + 9.124 * 10^{-2}T - 8.635 * 10^{-5}T^2 + 4.1382 * 10^{-8}T^3 \tag{4}$$

Similarly, a relation for Si can be obtained from the work by *Li*[2] as it provides an expression for *n(T,λ)* for a large temperature range of 20K-1600K

However, Eq. (2) does not take into account the effect of bandgap shrinkage with temperature, on the concentration of thermally generated free carriers (FC). In general, the FC contribution to the semiconductor permittivity is described by a simple Drude model[1]:

$$\varepsilon = \varepsilon_\infty - \frac{\omega_p^2}{\omega^2 + i\gamma\omega} \tag{5}$$

where the plasma frequency $\omega_p$, the plasma wavelength $\lambda_p$ the damping coefficient and scattering time τ are defined as $\omega_p = \frac{2\pi c}{\lambda_p} = \sqrt{\frac{ne^2}{m_c \varepsilon_0}}$ and $\frac{1}{\gamma} = \tau = \frac{\mu m_c}{e}$, $n$ is the free carrier concentration (not to be confused with the refractive index n), $e$ is the electron charge, $m_c$ is the conductivity effective mass, $\varepsilon_0$ and $\varepsilon_\infty$ are the permittivity of free space and the high frequency permittivity, respectively, and μ is the free carrier mobility. In the current case, the plasma frequency $\omega_p$ is temperature dependent as it depends on the carrier density (which is temperature dependent) $\omega_p$=f(T). Also, the high frequency permittivity $\varepsilon_\infty$ is temperature dependent: $\varepsilon_\infty$= $\varepsilon_{\infty,TO}$(T), and is obtained from the normal TOE on permittivity and refractive index as described in Eq. (1) and Eq. (2). Neglecting all other secondary effects of temperature (such as effective mass, mobility, band structure changes etc'), the explicit form of the temperature dependent permittivity now reads:

$$\varepsilon(T) = n(T)^2 = \varepsilon_{\infty,TO}(T) - \frac{\omega_p^2(T)}{\omega^2 + i\gamma\omega} \tag{6}$$

where the first term on the right hand side, $\varepsilon_{\infty,TO}$, is responsible for the normal TOE with no FC effects, and the second term for the FC contribution. Using Eq (6) we can calculate $n(T)$, $\frac{dn}{dT}(T)$ and its corresponding $(dn/dT)_{TO}$ and $(dn/dT)_{FC}$ components which contributes to the *total dn/dT* as:

$$\frac{dn}{dT} = \left(\frac{dn}{dT}\right)_{TO} + \left(\frac{dn}{dT}\right)_{FC} \quad (7)$$

where the first term is Eq. (7) is the normal TO coefficient as depicted in Eq. 2 and the second term refers to FC effects on the refractive index. This effect of FCs cannot be neglected at elevated temperatures and is more prominent for semiconductors with narrower bandgaps. The temperature dependence of energy bandgap in Si and Ge can be expressed as[3]:

$$E_g(T) = E_g(0) - \frac{\alpha T^2}{T+\beta} \quad (8)$$

where $E_g(0)$, α and β are material dependent fitting parameters. Figure S1 presents the energy bandgap as a function of temperature for both Si and Ge, with the fitting parameters taken from ref [3]. The intrinsic carrier density can be expressed as a function of density of states in the conduction and valence bands[3]:

$$n_i = \sqrt{N_c N_v} e^{-E_g/2kT} \quad (9)$$

where the effective density of states in conduction and valence bands is given by[3]:

$$N_c = 2\left[\frac{2\pi m_e^* kT}{h^2}\right]^{3/2} \quad (10)$$

$$N_v = 2\left[\frac{2\pi m_h^* kT}{h^2}\right]^{3/2} \quad (11)$$

with $m_e$ and $m_h$ being the effective masses of the electron in hole in conduction and valence band, respectively. Equations (8), (10) and (11) were used to calculate the intrinsic carrier density $n_i(T)$ in Eq. (9) and are presented in Figure S2. The carrier density $n_i(T)$ was then used to obtain $\omega_p(T)$ which together with the known values for $\varepsilon_{\infty,TO}(T)$ (taken from ref [1] and [2] as mentioned above) allows to calculate the full expression for n(T), Δn(T) and dn/dT using Eq. (6) and Eq. (7).

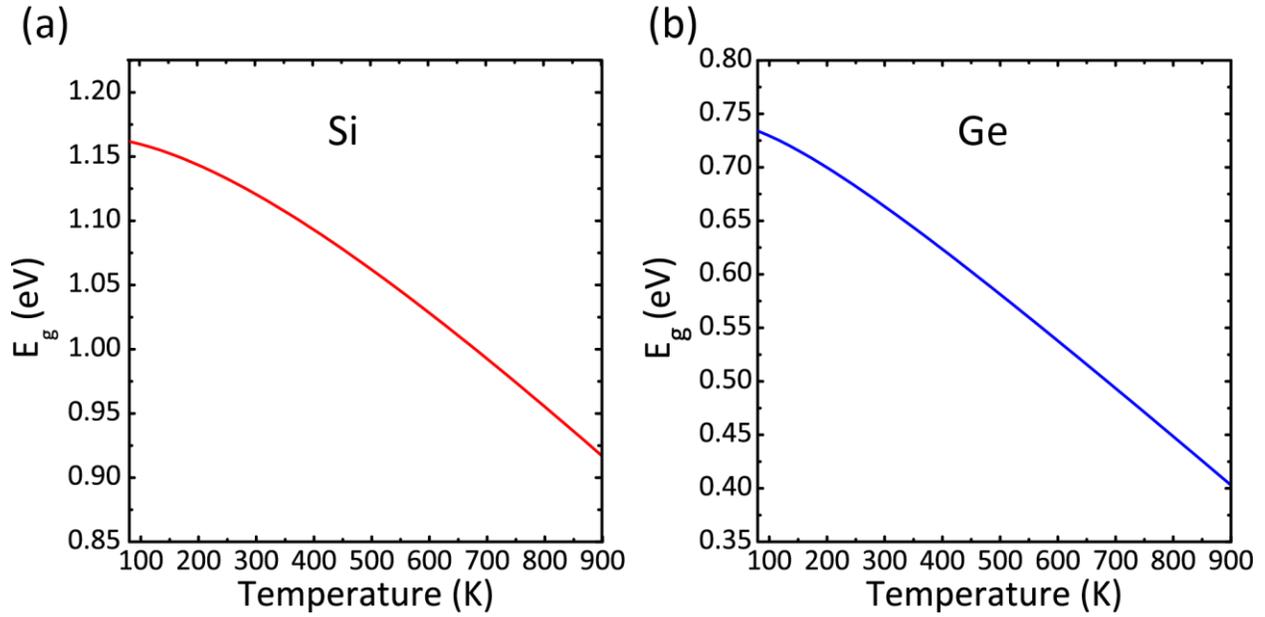

Figure S1: Temperature dependence of the Energy bandgap in (a) Si and (b) Ge

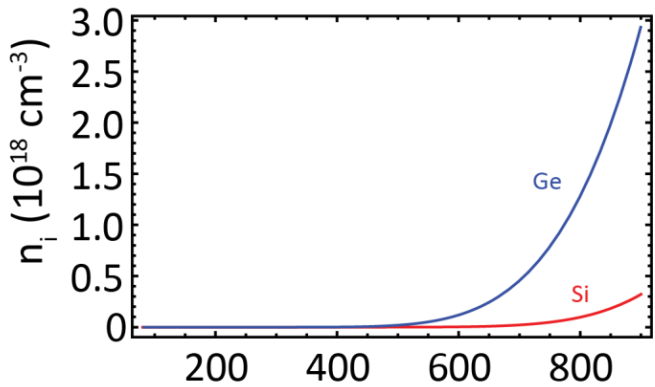

Figure S2: Intrinsic carrier density in Si and Ge as a function of temperature. The higher carrier density of Ge compared to Si at a given temperature, is due to its lower effective mass and narrower bandgap.

## 2. Comparison of calculated and measured Spectra

Incorporating our modified Drude model, which now takes into account the generation of free carriers along with the traditional thermo-optic effect, we generate scattering cross-section calculations for the two Silicon and Germanium spherical resonators (which are discussed in the main text) and compare them to the measured reflection spectra. As an example, Figure S3 presents the calculated and measured spectra of the Si (r=1.9µm) and Ge (r=1.54µm) spherical Mie resonators at T=573K and T=773K, respectively. At both temperatures, experimental spectra show good agreement with the calculated Mie scattering normalized to the geometric cross-section $\sigma_{geo}$. Naturally, it is consistent with the extracted induced index and resonance wavelength shifts presented in Figures 2(e) and 2(f) of the main text, and validates of our temperature dependent refractive index model.

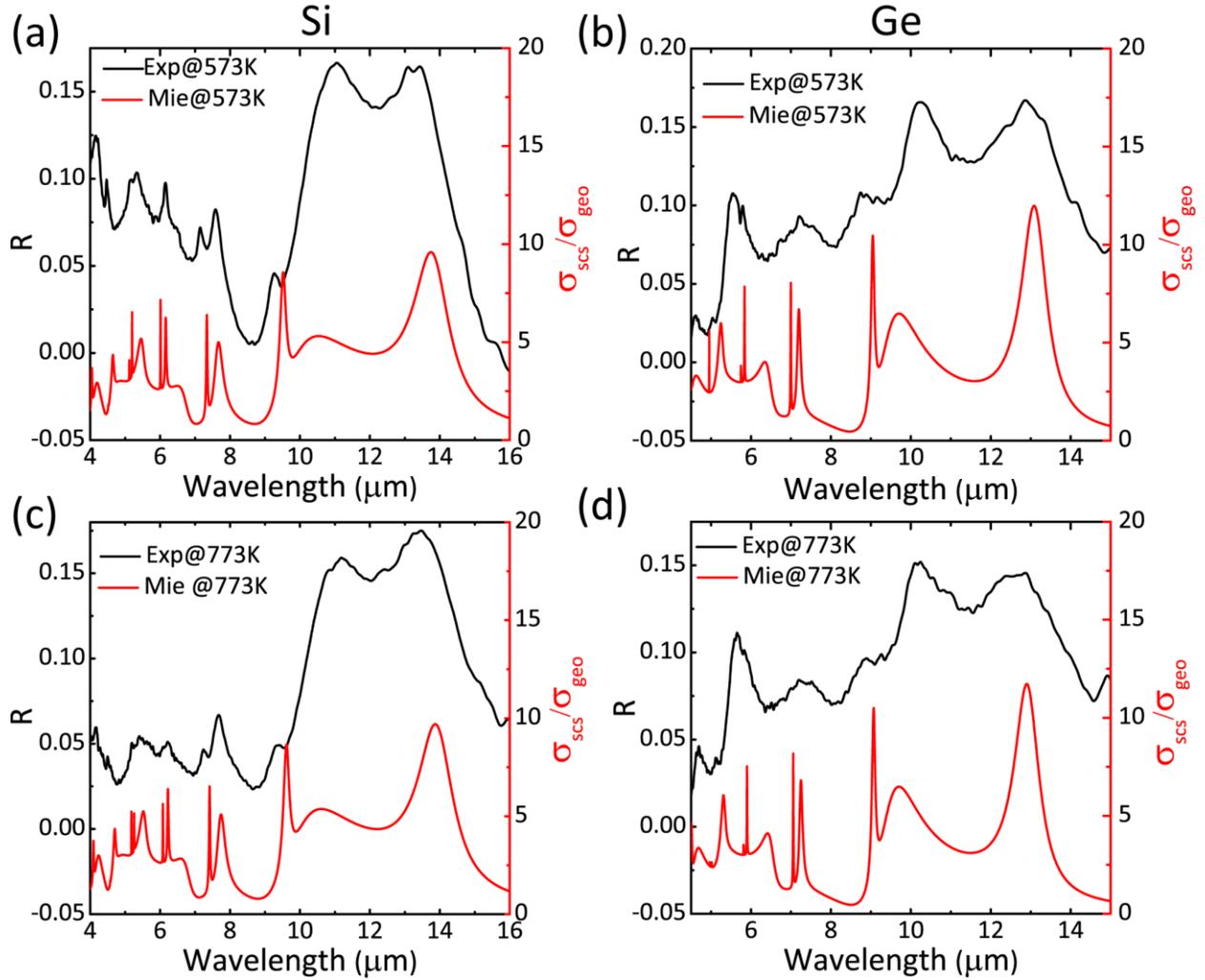

Figure S3: Measured and calculated infrared spectra of Si ((a) & (c); r=1.9µm) and Ge ((b) & (d); r=1.54µm) spherical Mie resonators at T=573K and T=773K, respectively. Experimental spectra show good agreement with the calculated Mie scattering normalized to the geometric cross-section $\sigma_{geo}$.

### 3. Extracted induced index shift Δn(T)

In order to extract the induced index shift, Δn(T), from experimental measurements we used the following procedure:

The effective refractive index at a particular resonance is obtained from the resonance condition derived from Mie theory[4]: $2\pi rn/\lambda$=constant, where r is the particle radius, n is the refractive index and λ is the free space resonance wavelength. Measuring the particle radius and resonance wavelength allows to extract the refractive index. By tracking the resonance wavelength shift of each resonance, as a function of temperature, we can deduce the corresponding index change. The induced index shift Δn(T) for each resonance is calculated relative to the index at RT, that is, Δn(T)=n(T)-n(T=293K).